\documentclass[summary]{ursi}
\usepackage{xcolor}
\usepackage{hyperref}
\usepackage{caption}
\usepackage{subfig}
\usepackage{graphicx}

\title{Reconfigurable Intelligent Surface-assisted Classification of Modulations using Deep Learning}

\author{Mir~Lodro\affref{ref1}, Hamidreza~Taghvaee\affref{ref1}, Jean-Baptiste~Gros \affref{ref2}, Steve~Greedy\affref{ref1} and Geofrroy~Lerosey\affref{ref2}
  and Gabriele~Gradoni\affref{ref1}\affref{ref3}}

\affiliation{%
  \aff{ref1}{George Green Institute for Electromagnetics Research-GGIEMR, University of Nottingham, UK}
  \aff{ref2}{Greenerwave, Paris, France}
  \aff{ref3}{Maxwell Centre, Cavendish Laboratory, University of Cambridge, UK}
  
}

\begin{document}

\maketitle

\begin{abstract}
The fifth generating (5G) of wireless networks will be more adaptive and heterogeneous. Reconfigurable intelligent surface technology enables the 5G to work on multistrand waveforms. However, in such a dynamic network, the identification of specific modulation types is of paramount importance. We present a RIS-assisted digital classification method based on artificial intelligence. We train a convolutional neural network to classify digital modulations. The proposed method operates and learns features directly on the received signal without feature extraction. The features learned by the convolutional neural network are presented and analyzed. Furthermore, the robust features of the received signals at a specific SNR range are studied. The accuracy of the proposed classification method is found to be remarkable, particularly for low levels of SNR.
\end{abstract}

\section{Introduction}

The reconfigurable intelligent surface (RIS) is a new promising technology for 5G wireless communication, sensing and localization \cite{strinati2021wireless}\cite{basar2019wireless}\cite{wymeersch2020radio}\cite{alexandropoulos2021reconfigurable}. Because of its superior control on the impinging electromagnetic (EM) waves, it can create a reconfigurable smart EM environment \cite{di2020smart}. The deployment of multiple RIS devices for communication applications is attractive because of their key benefits such as: i) use passive components; ii) absence of dedicated energy sources. RISs can be a continuous surface (aperture) or a discrete surface (reflectarray or metasurface). Additionally, the RIS has the ability to tune the operational frequency within its resonance bandwidth. RISs can be easily deployed by installation over building facades, within rooms, shopping malls, and sports arena. Since the RIS unit cell does not require the integration of components like analog-to-digital converters (ADCs) and amplifiers, the energy-consumption of the RIS is low. Furthermore, the RIS is considered as quasi-passive as no RF power amplification is required for their dynamic operation, besides DC power needed to operate tunale electronics components, e.g., p.i.n. diodes and varactors. 
These features make RISs competitive against other communication technologies, e.g., amplify and forward relaying and massive MIMO. The advantages of  the RIS compared to relaying technologies have been addressed by some investogators. For instance, in \cite{boulogeorgos2020performance} RIS-assisted communication are compared with passive relays. The Authors present end-to-end expressions for SNR, outage probability, symbol error rate, and the ergodic capacity of relaying and the RIS-assisted links in fading channels. Since RIS consists of a large number of cost-effective and passive elements, it offers improved energy-efficiency \cite{huang2019reconfigurable}, spectral-efficiency, coverage enhancement, and PHY layer security. The RIS alleviates the negative effect of multipath-fading and can benefit from channel hardening \cite{bjornson2020rayleigh} for robust wireless communication system operation. To date experimental setups have been demonstrated to show the benefits of RIS integration in active Tx-Rx links in multimode metallic enclosures \cite{lodro2021reconfigurable},indoor \cite{dai2020reconfigurable}, as well as in outdoor environments \cite{pei2021ris}. The Authors in \cite{zhou2021modeling} have shown experimental setups with signal generator, RIS and directional antennas. The purpose of the setup was to perform multipath mitigation using RIS. Although the RIS technology addresses key issues in wireless communication systems, the idea of RIS-assisted modulation classification remains unexplored in the literature. Modulation classification are widely used in the military and commercial applications. Automatic modulation classification plays a key role in 5G cellular communication and beyond, mainly for  efficient radio resource sharing and management \cite{kaleem2021artificial}. The experimental setup for the modulation classification can also be useful for the detection of video and control links of rogue drones that are threat to a secure airspace. This work demonstrates the benefit of integration of RIS for automatic modulation classification. The rest of the paper is arranged as follows. Section II describe RIS-assisted experimental setup. Section III explains indoor propagation environment where the measurements are conducted. Section IV describes the data set generation and the training of a convolutional neural network (CNN). Section V is dedicated to quantifying digital modulation classification accuracy.
\section{Experimental Setup}
\begin{figure}
    \centering
    \includegraphics[width=\columnwidth]{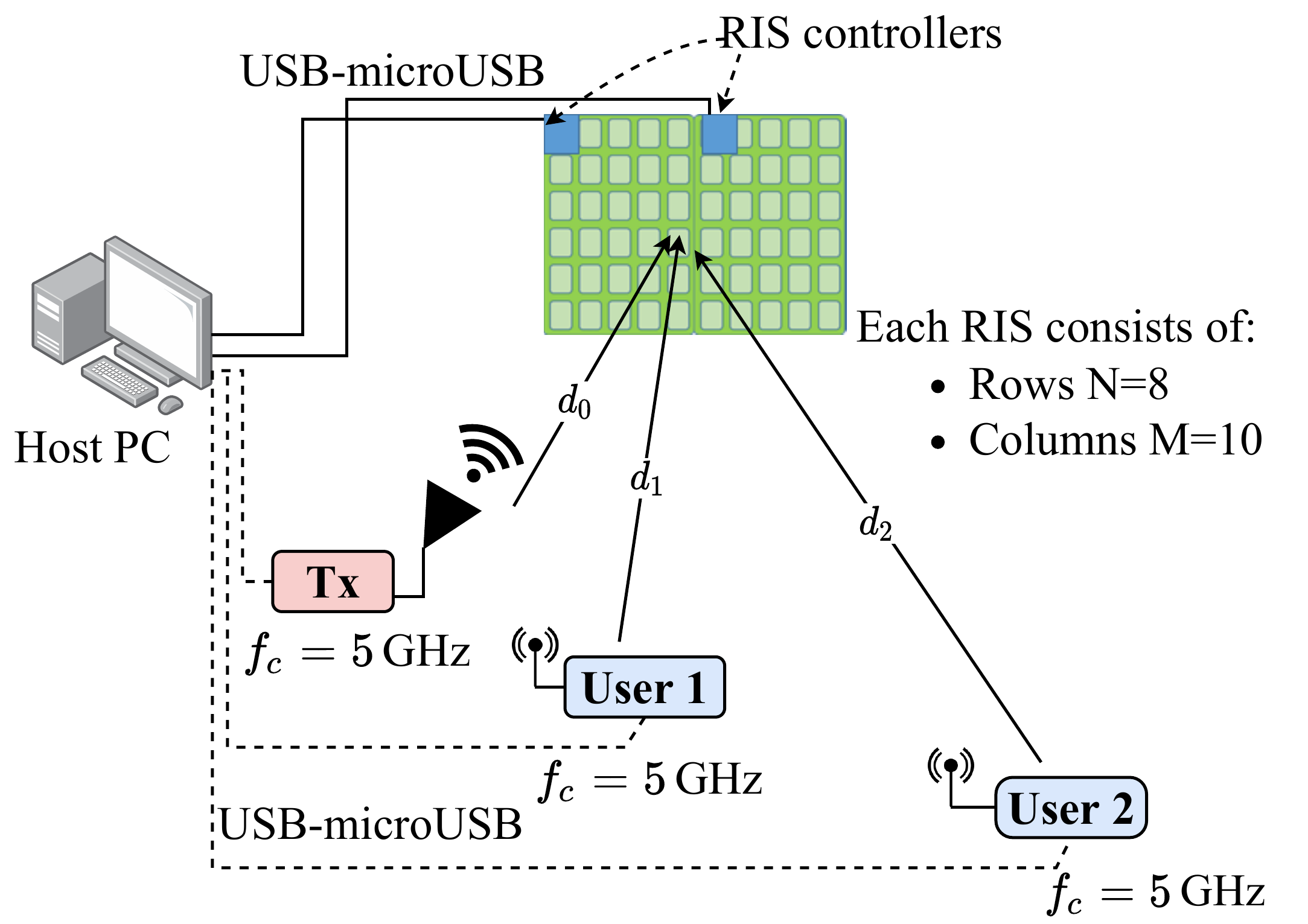}
    \caption{Diagram of multi-user RIS-assisted links.}
    \label{fig:sketch}
\end{figure}

The multi-user links that need optimization are created using RISs and PlutoSDR from Analog Devices. PlutoSDR is a single channel software-defined-radio that supports full-duplex transmit-receive operations. The multi-user RIS-assisted experimental setup consists of two RISs and three PlutoSDRs that are connected to the same host PC using micro-USB 2.0 cable. One PlutoSDR is configured as a transmitter that generates real-time complex samples of all the digital modulations, which are up-converted for over-the-air transmission using log-periodic antenna. The transmitter uses LP09650 850 MHz to 6.5 GHz Log-Periodic PCB directional antenna with 5-6 dBi gain and is directed towards the center of two RISs. The measurement set-up is shown in Fig.\ref{fig:sketch}. The two PlutoSDRs configured as user 1 and user 2 adopt identical monopole antennas for signal reception. We deployed two RISs from GREENERWAVE \mbox{(\url{https://greenerwave.com/})} between transmitter and the user 1 and user 2. The RISs are connected to the host PC using a USB port. The two RISs get power, command, and control signals via USB. The RISs configuration is changed by controlling the bias voltage of the PIN diodes through a set of shift registers. Each RIS consists of 76 phase-binary pixels \cite{kaina2014hybridized} and each pixel can be controlled independently via states of two PIN diodes, by imposing 0 or $\pi$ phase, thus obtaining a total number of 152 effective pixels per RIS. The two RISs are co-located and the transmitter and receivers are operated in a single-input-single-output (SISO) mode. There is no line-of-sight LOS between transmitter and the two users so that the received signals are generated by reflection off the two RISs. We denoted distance between center of the two RISs and the transmitter as $d_0$ and the distance between center of the RISs to user 1 as $d_1$ and the distance between center of the RISs and the user 2 as $d_2$.The azimuth ($\phi$) and elevation ($\theta$) angles for the measurement setup is such that the user 2 is located at $(0^{\circ},35^{\circ)}$, user 1 is at $(0^{\circ},120^{\circ})$ and the Tx is located at $(0^{\circ},110^{\circ})$ angles from the origin $(\theta=0^{\circ},\phi=0^{\circ})$ of the RIS. The multi-user RIS deployment is performed in the indoor office environment. The optimization of modulation classification accuracy are performed at RF frequency of 5 GHz. The RISs are connected to the same host PC. We generated shaped beams by optimizing the binary coding reflection matrices of the RISs. This increases the signal strength at two users which has direct implications to increased modulation classification test accuracy. The CNN is also trained to classify digital modulation techniques in the presence of hardware impairments and Rician fading with K-factor of 4. All the complex digital modulation samples used for training purpose were subject to channel impairments with maximum Doppler shift of 10 Hz, Clock offset of 5 parts per million (ppm). The CNN was trained to classify the samples at SNR of 10 dB. The measurement parameters for transmitter and the two users are shown in Table.\ref{tab:table_param}. The transmitter and the two PlutoSDRs configured as user 1 and user 2 were operated at sample rates of 200 KS/s. The sample size was 2048.
\section{Indoor Propagation Environment}
The indoor office environment consists of office desks, metal shelves, PCs and monitors for around 8 people. However, at the time of measurement there were no more than 4 people entering and leaving the room. The measurements were done at the 7th floor of a 15-story building with reinforced concrete walls. There was no LOS between Tx and the user 1 and user 2, respectively, and only the cascaded channels were optimized. The environment is quasi-stationary. The operating frequency of the measurement is 5 GHz. The RISs, the transmitting PlutoSDR and the receiving PlutoSDRs are placed at the same height of $1\,m$ from the ground. RF scanning of 5 GHz was done to know the presence of active transmissions around 5 GHz that may introduce interference to our measurement link. However, there were sparse transmissions and hence the optimization accuracy is done in the presence of interference if in case there were active channels.
\begin{table}
    \centering
     \caption{MEASUREMENT PARAMETERS}
    \begin{tabular}{|c|c|}
    \hline
    Parameter & Value \\
    \hline
    Tx PlutoSDR Gain     &  0 dB\\
    Rx PlutoSDR1 Gain     &  45 dB\\
    Rx PlutoSDR2 Gain & 62 dB\\
   Center frequency $f_c$       & 5 GHz\\
    Sample rate & 200 KS/s\\
    \hline
    \end{tabular}
    \label{tab:table_param}
\end{table}
\section{Dataset and CNN Training}
\begin{figure}
    \centering
    \subfloat[Complex Samples]{\includegraphics[width=\columnwidth]{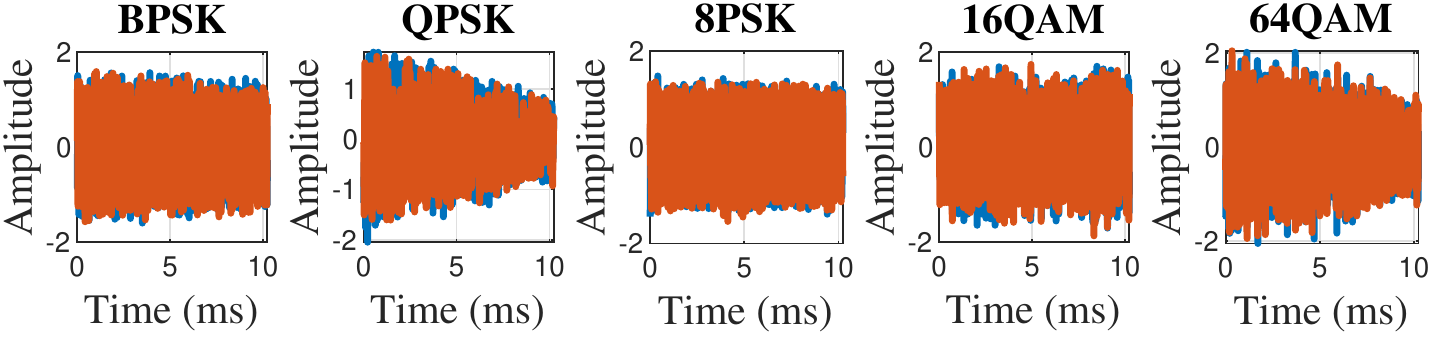}}\\
    \subfloat[Spectrogram of the complex samples]{\includegraphics[width=\columnwidth]{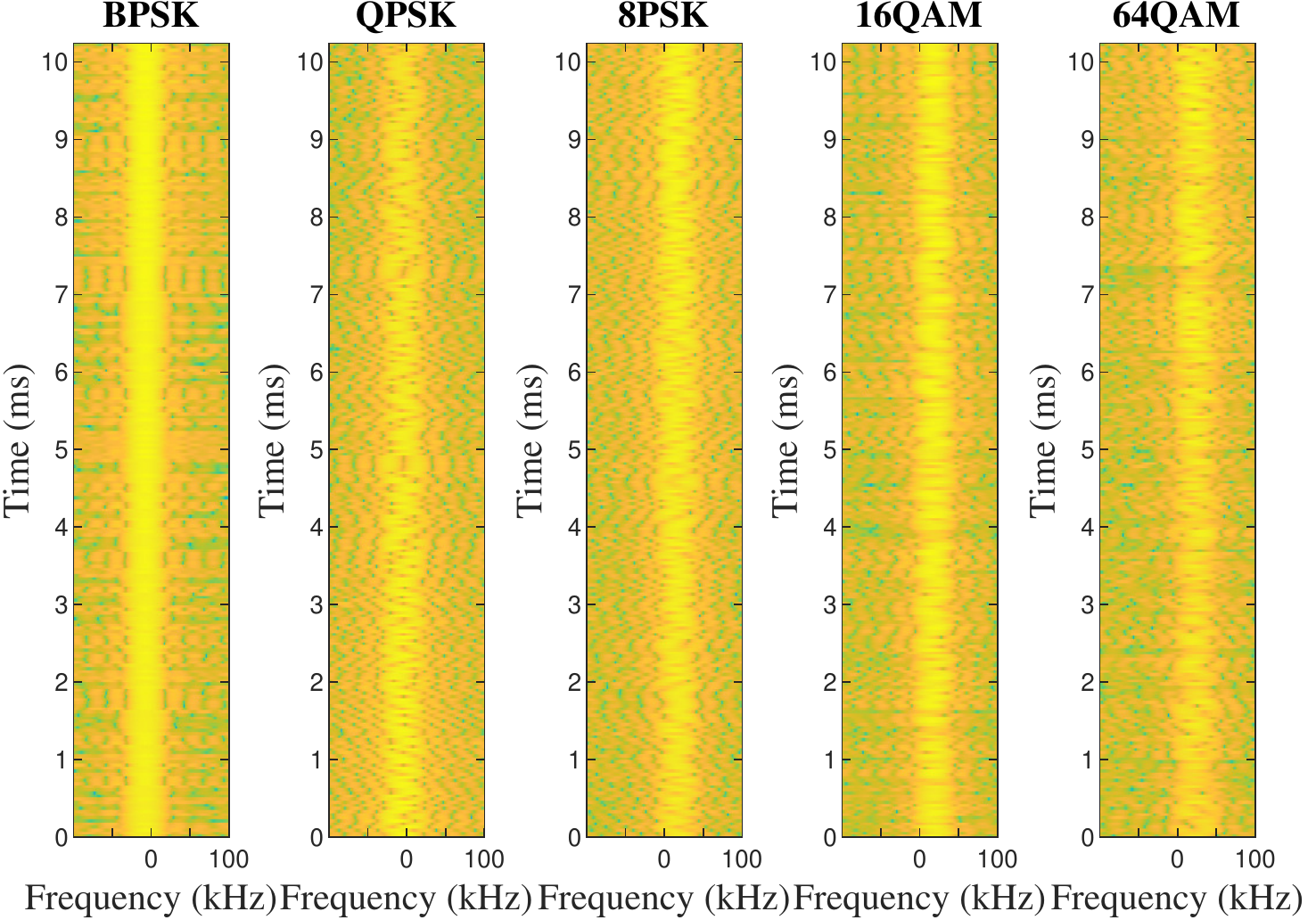}}
    \caption{Complex samples and the spectrogram of the complex samples of digital modulation techniques.}
    \label{fig:complex_samples}
\end{figure}

The dataset includes five digital modulations that are used in modern communication systems: BPSK, QPSK, 8-PSK, 16-QAM and 64-QAM. The complex samples of each of the five digital modulation techniques are generated, which are subject to channel and hardware impairments. The complex samples are generated in the presence of frequency offset, timing offset, AWGN and Rician fading with K-factor of 4. Figure \ref{fig:complex_samples} shows complex samples of digital modulation techniques and the corresponding spectrogram. Out of five digital modulation techniques three are phase-based whereas two modulation techniques, the 16-QAM and 64-QAM, involve both amplitude and phase modulation. The trained CNN consists of six convolution layers and one fully connected layer. Each convolution layer is followed by batch normalization layer, rectified linear unit (ReLU) activation layer, and max pooling layer. Stochastic gradient descent with momentum (SGDM) solver is used with mini-batch size of 256. Initial learning-rate is set to 0.02 and it is reduced by a factor of 10 every 9 epochs. We used 5000 frames for each of the digital modulation techniques for training purpose where $80\,\%$ is used for the training, $10\,\%$ is used for validation and $10\,\%$ is used for the testing purpose. Each frame of the digital modulation techniques is 2048 long and generated at sample rate of 200 KS/s. The same sampling rate of 200 KS/s is used for the PlutoSDRs configured as transmitter, user 1 and user 2 respectively. The CNN is trained to take 2048 channel impaired samples and predict modulation type of each frame, and measure the test accuracy of each modulation technique. The training progress of the CNN was monitored, which shows accuracy and loss of the training process. The CNN converges in 12 epochs with validation of $99.80\,\%$. The number of epochs per iteration are 78. The training progress takes around 20 minutes on Intel Xeon PC with a small form-factor NVIDIA Quadro P400 GPU. The GPU has memory of 2 GDDR5, and it has memory bandwidth up to 2 GB/s.   

\section{Measurement Results}
\begin{figure}[htbp]
    \centering
    \subfloat[$54\%$]{\includegraphics[width=0.5\columnwidth]{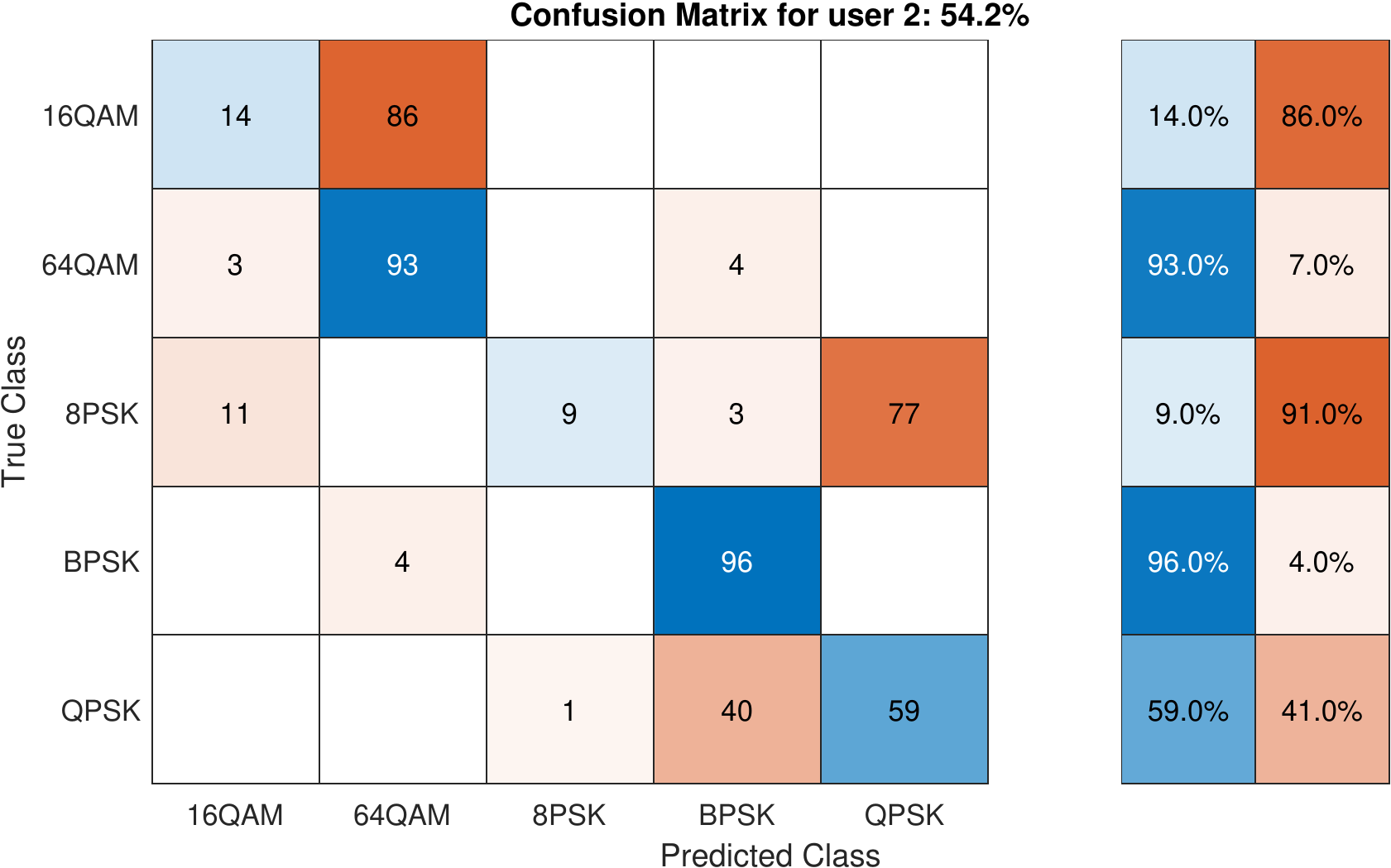}}
    \subfloat[$98\%$]{\includegraphics[width=0.5\columnwidth]{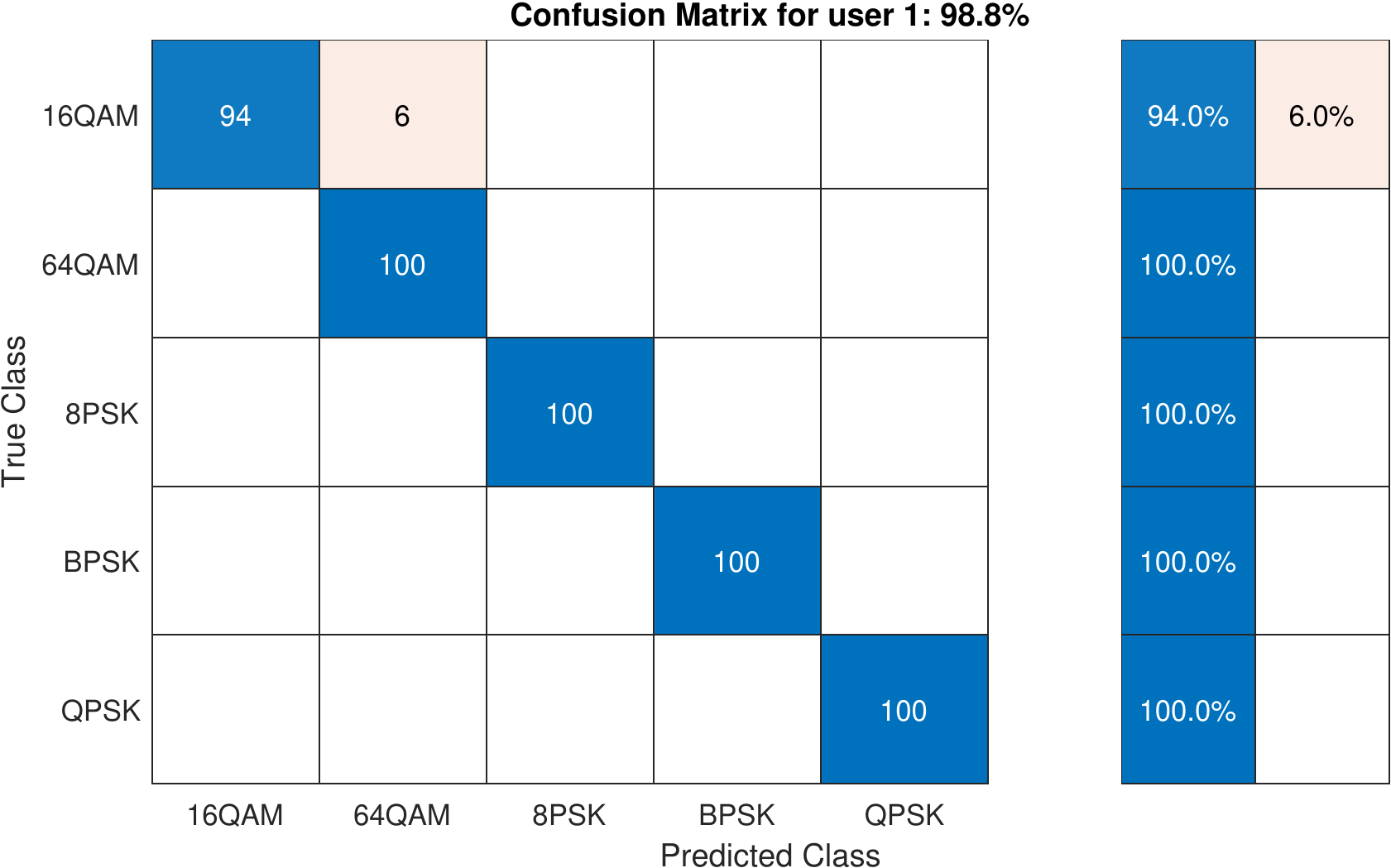}}\\
    \subfloat[$96\%$]{\includegraphics[width=0.5\columnwidth]{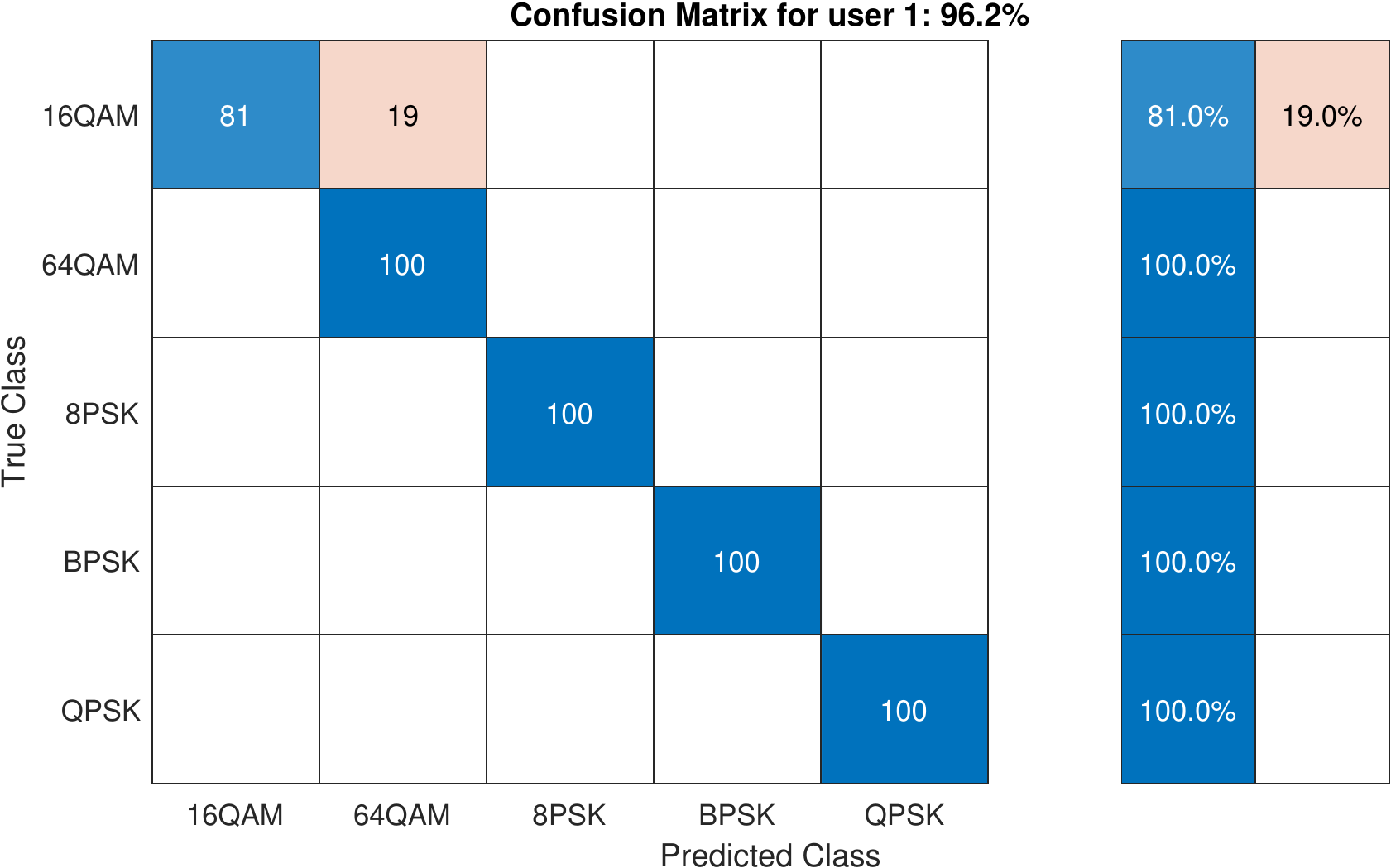}}
    \subfloat[$99\%$]{\includegraphics[width=0.5\columnwidth]{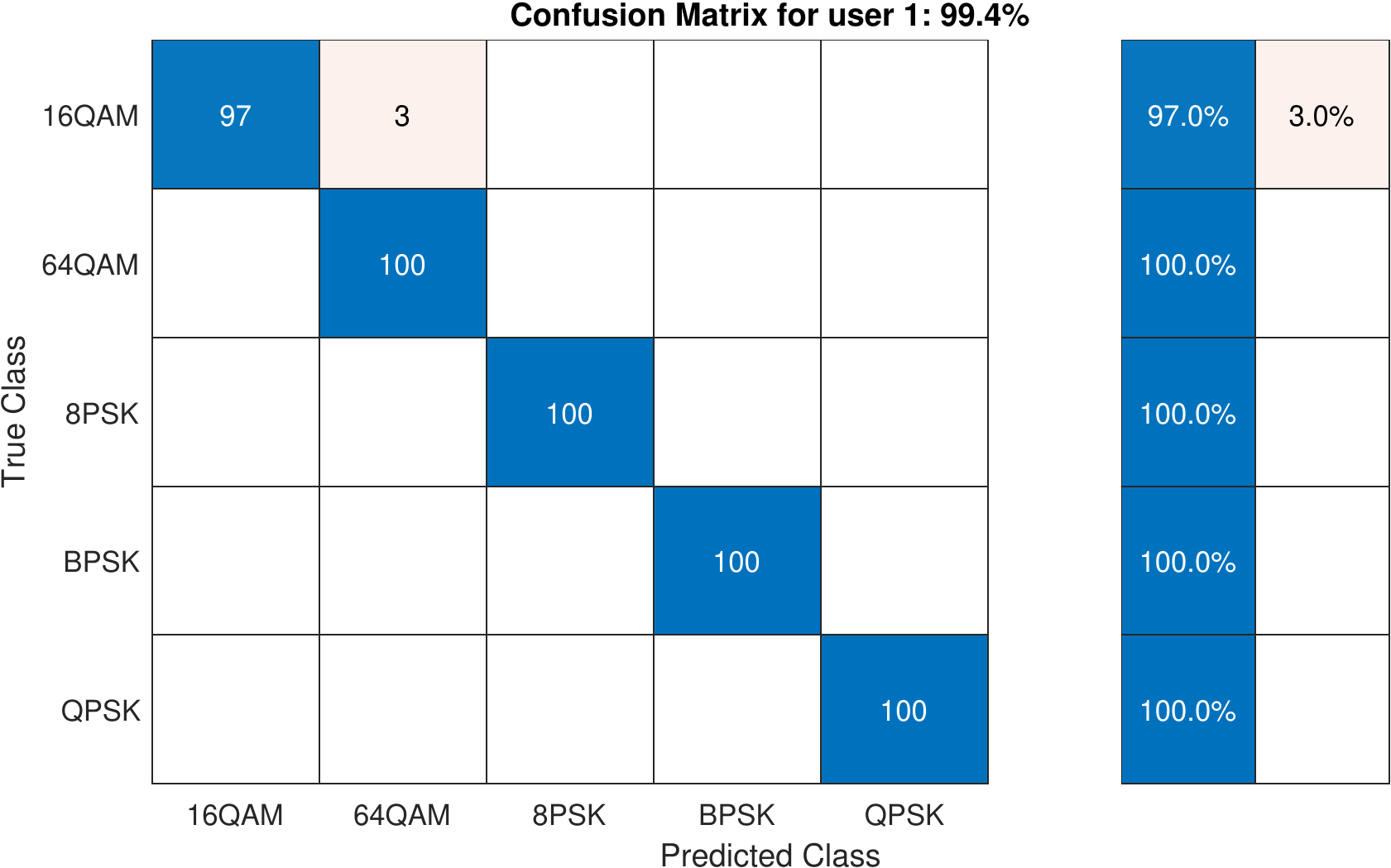}}
    \caption{Confusion matrix for modulation classification for user 1 and user 2. }
    \label{fig:conf_matrix}
\end{figure}

\begin{figure}[htbp]
    \centering
    \includegraphics[width=\columnwidth]{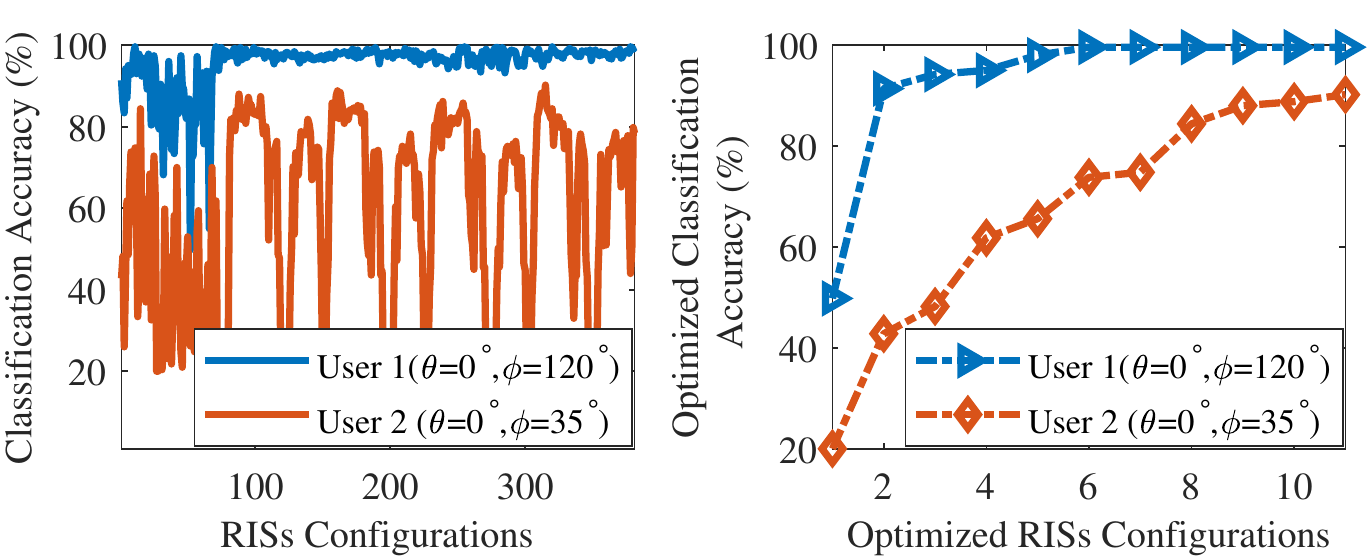}
    \caption{Instantaneous accuracy and the corresponding optimized modulation classification accuracy of user1 and user 2 respectively.}
    \label{fig:opt_acc_10}
\end{figure}

In order to observe the overall modulation classification test accuracy, the ground truth and the predicted class of each modulation is plotted side by side in a confusion matrix. The total number of 100 frames for each modulation technique are used for the validation of the test. The total elapsed time to get accurate  RIS-optimized classification of modulations was around 20.6 hours. The total time includes the transmission and detection times of each of the digital modulation technique per RIS configuration. Figure \ref{fig:conf_matrix} shows the confusion matrix of user 1 and user 2 respectively, we observe that there is classification inaccuracy for 16-QAM and 64-QAM as the 16-QAM is subset of 64-QAM. Additionally, there is also classification inaccuracy between QPSK and 8PSK digital modulation techniques. The reason for classification inaccuracy of QPSK is the presence of frequency offset. In presence of frequency offset, the QPSK constellation looks like a subset of 8PSK. Additional inaccuracy factors can be noisy samples in the presence of frequency offset and clock offset. However, the overall classification accuracy is improved by optimizing the phase shift matrix of the two RISs. There is less classification inaccuracy for BPSK as the two constellation points are separated by a decision boundary with less chance of errors. During the measurements the chance of less classification inaccuracy for BPSK was observed for both the users. During all the experiments two confusion factors were predominant: i) classifier confusing 16-QAM with 64-QAM; ii) classifier confusing QPSK with 8-PSK. Figure \ref{fig:opt_acc_10} shows theinstantaneous modulation classification accuracy and the corresponding optimized modulation classification accuracy for user 1 and user 2. It can be seen that the RIS introduction increases modulation classification accuracy from $20\%$ to $100\%$ for user 1, which is a $400\%$ increase in the modulation classification accuracy. Similarly, the modulation classification accuracy has increased from $20\%$ to $90.2\%$ for user 2, which is a $351\%$ increase in the modulation classification accuracy. Additionally, we can see that there are four RISs configurations where we can get modulation accuracy greater than $80\%$ for both the users.

\section{Conclusion}
The RIS-assisted classification accuracy for two users at different distances from the RISs has been improved in an indoor office environment. The optimization of final test accuracy of modulation classification reflects that the RISs can be used to create a virtual LOS that can be used to increase the SNRs at two users. The RISs have been used for passive beamforming targeted for two users located at different positions in indoor propagation environment. We found that the RISs-assisted optimized link can increase the modulation classification accuracy and the trained CNN can fully classify modulated signal for user 1 and it can achieve modulation classification accuracy upto $90.2\%$ for user 2. The work can be extended to a higher number of users by adopting omnidirectional transmitting antenna, thus fully exploit the degree of freedom introduced by the RISs.
\section{Acknowledgements}
This work has been supported by the European Commission through the H2020 RISE-6G Project under Grant 101017011 and InnovateUK project no. 71067 "Edged Sensing Array Affording Intelligent Integrated Airspace Awareness". The work of Gabriele Gradoni was supported by the Royal Society under Grant INF$\backslash$R2$\backslash$192066.

\bibliographystyle{IEEEtran}

\bibliography{mod_class.bib} 


\end{document}